\batchmode
\documentclass[12pt,a4paper,ngerman,english]{iopart}
\usepackage[T1]{fontenc}
\usepackage[latin9]{inputenc}
\usepackage{amstext}
\usepackage{graphicx}
\usepackage{amssymb}

\makeatletter
\usepackage{iopams}
\usepackage{setstack}

\usepackage{verbatim}

\makeatother

\usepackage{babel}

\begin{document}

\title{Inferring Network Topology from Complex Dynamics}

\author{Srinivas Gorur Shandilya$^{\text{1}}$ and Marc Timme$^{\text{1,2,3}}$}

\address{$^{\text{1}}$Network Dynamics Group, Max Planck Institute for Dynamics
\& Self-Organization, 37073 Göttingen, Germany,}

\address{$^{\text{}2}$Bernstein Center for Computational Neuroscience, 37073
Göttingen, Germany, Germany,}

\address{$^{3}$Faculty of Physics, University of Göttingen, 37077 Göttingen,
Germany.}

\ead{timme@nld.ds.mpg.de}
\begin{abstract}
Inferring network topology from dynamical observations is a fundamental
problem pervading research on complex systems. Here, we present a
simple, direct method to infer the structural connection topology
of a network, given an observation of one collective dynamical trajectory.
The general theoretical framework is applicable to arbitrary network
dynamical systems described by ordinary differential equations. No
interference (external driving) is required and the type of dynamics
is not restricted in any way. In particular, the observed dynamics
may be arbitrarily complex; stationary, invariant or transient; synchronous
or asynchronous and chaotic or periodic. Presupposing a knowledge
of the functional form of the dynamical units and of the coupling
functions between them, we present an analytical solution to the inverse
problem of finding the network topology. Robust reconstruction is
achieved in any sufficiently long generic observation of the system.
We extend our method to simultaneously reconstruct both the entire
network topology and all parameters appearing linear in the system's
equations of motion. Reconstruction of network topology and system
parameters is viable even in the presence of substantial external
noise. 
\end{abstract}


\maketitle

\section{Background}

Understanding the relations between network topology and its collective
dynamics is at the heart of current interdisciplinary research on
networked systems \cite{Strogatz:2001p9255}. Often it is possible
to observe the dynamics of the individual units of the network, whereas
the coupling strengths between them and the underlying network topology
cannot be directly measured. Hence, various methods have been proposed
to solve the inverse problem of inferring network structure from observation
and control of dynamics.

Perturbing a fixed point of a network dynamical system constitutes
the simplest controlled intervention of a system. The method of Tegnér
et al.~\cite{Tegner:2003p9754} perturbs the steady state expression
levels of selected genes. Their iterative algorithm can reveal the
structure of an underlying gene regulatory network by analysing resultant
dynamical changes in the pattern of gene expression levels. A similar
iterative method based on multiple regression, coupled with transcriptional
perturbations to the fixed points of a genetic network, has been used
to successfully identify a nine-gene sub-network~\cite{Gardner:2003p9614}.
A method introduced by Timme~\cite{Timme:2007p14319} extends reconstruction
to networks of smoothly coupled limit-cycle oscillators with periodic
collective dynamics. The underlying idea is that the asymptotic response
dynamics of a network to different externally induced driving conditions
is a function of its topology \emph{and} of the (external) driving
signals. Thus measuring the response to suitable driving signals in
different experiments restricts the set of network topologies that
are consistent with the driving-response pairs, yielding the network's
topology for sufficiently many experiments.

Can we reconstruct a network displaying collective dynamics richer
than simple fixed points or limit-cycles? Yu et al.~\cite{Yu:2006p13046}
introduced a synchronization method to identify networks of chaotic
Lorenz oscillators up to \emph{N}=17 units. Assuming full knowledge
of all model parameters, the network topology of a clone of the system
is varied progressively via error minimisation until it synchronizes
with the original system. The topology of the clone is then recognised
as that of the original network. An extension of this synchronisation
method~\cite{Yu:2008p13684} involves additional `control signals'
to externally drive the system to steady states, allowing the inference
of interaction topology for sparse, symmetric networks.

Here, we introduce a simple, direct and intervention-free method to
reconstruct networks of arbitrary topology from the mere observation
of generic collective dynamics. Given the functional form of the intrinsic
and interaction dynamics, we show that all other factors of the network
dynamical system, such as network topology and coupling strengths
(and even typical parameters), can be reliably and efficiently reconstructed.

\section{Theory of Direct Reconstruction from Dynamical Trajectories}

\begin{figure}[t]
\hspace{20mm}\includegraphics[clip,width=0.75\columnwidth]{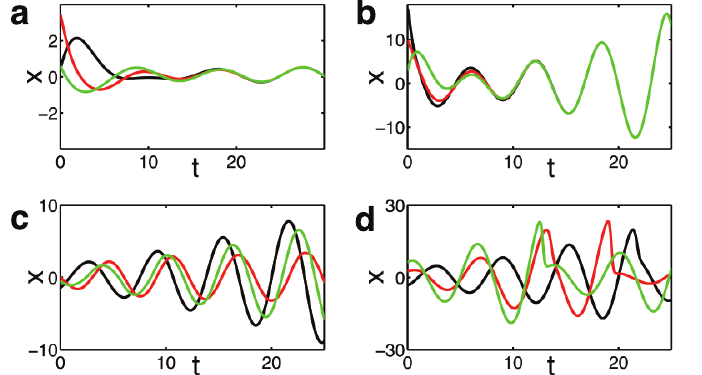}

\caption{\textsf{\textbf{\small \label{fig:fig1}}}Reconstruction of networks
is possible for various types of collective dynamics. A 32-unit random
network of Rössler oscillators (\ref{eq:rossler-eq}) with fixed connection
probability $p=0.5$ exhibiting (a) a synchronous periodic state ($a_{i}\equiv0.2$,
\emph{$b_{i}\equiv1.7$} and $c_{i}\equiv4.0$ for all units); (b)
a synchronous chaotic state ($a_{i}\equiv0.2$, \emph{$b_{i}\equiv1.7$}
and $c_{i}\equiv13.0$ for all units); (c) an asynchronous periodic
state (parameters randomly drawn independently from $a_{i}\in[0.10,0.101]$,
$b_{i}\in[0.10,0.101]$ and $c_{i}\in\{4,6,12\}$); (d) an asynchronous
chaotic state (parameters randomly drawn independently from $a_{i}\in[0.0,0.38]$,
$b_{i}\in[0.0,2.0]$ and $c_{i}\in[13,16]$) The \emph{x}-values of
three out of $N=32$ oscillators are shown. Close to perfect reconstructions
from all these distinct dynamics are shown in Fig.~\ref{fig:reconstruction}.}

\end{figure}

Given an observation of the collective trajectory of a dynamical network,
how can we infer its underlying topology? We consider a dynamical
system consisting of \emph{N} units, where the dynamics of each unit
is specified by an arbitrary set of dynamical equations, and the interactions
between the units take place via edges in the network. The units of
the dynamical system are coupled on a directed graph of unknown connectivity
with their dynamics satisfying 

\begin{equation}
\frac{d}{dt}\mathbf{x}_{i}=\mathbf{f}_{i}\left(\mathbf{x}_{i}\right)+\sum_{j=1}^{N}J_{ij}\mathbf{g}_{ij}\left(\mathbf{x\mathit{_{i}}\textrm{,}}\mathbf{x}_{j}\right)\label{eq:general-equation}\end{equation}
where $i,j\in\{1,2,\ldots N\},$ and $\mathbf{x_{i}}=\left[x_{i}^{(1)},x_{i}^{(2)},\cdots,x_{i}^{(D)}\right]\in\mathbb{R^{\textrm{D}}}$
describes the state of the \emph{i}-th unit, and the functions $\mathbf{f}_{i},\mathbf{g}_{i}:\,\mathbb{R}^{D}\rightarrow\mathbb{R}^{D}$
mediate intrinsic and interaction dynamics of the \emph{D}-dimensional
unit, and are known. Methods based on copy-synchronisation~\cite{Yu:2006p13046,Yu:2008p13684}
rely on the construction of a new network, with dynamics governed
by Eq.~\ref{eq:general-equation} and network parameters $J'_{ij}$
that are tuned to that of the real network by an error minimisation
procedure. Here, we reduce the same reconstruction problem by evaluating
the states and their derivatives directly, recognising that the only
remaining unknowns in Eq.~\ref{eq:general-equation} are the coupling
strengths, which are to be determined. 

The dynamics of the \emph{d}-th dimension of the \emph{i}-th unit
is given by 

\begin{equation}
\dot{x}_{i}^{\left(d\right)}\left(t_{m}\right)=\mathit{f_{i}^{\left(d\right)}\left(\mathbf{x}_{i}\left(t_{m}\right)\right)+\sum_{j=1}^{N}J_{ij}g_{ij}^{\left(d\right)}\left(\mathit{\mathbf{x}_{i}\left(t_{m}\right),\mathbf{x}}_{j}\left(t_{m}\right)\right)}\label{eq:general-equation-2}\end{equation}
where $t_{m}\in\mathbb{R}$ are the times we evaluate Equation~\ref{eq:general-equation}
and we now write $\dot{z}$ for the rate of change $\frac{d}{dt}z$
of a scalar variable $z$. If there are \emph{$M$} such times, $m\in\left\{ 1,\ldots,M\right\} $,
we have \emph{M} equations of the form

\begin{equation}
\dot{x}_{i,m}^{(d)}=f_{i,m}^{(d)}+\sum_{j=1}^{N}J_{_{ij}}g_{ijm}^{(d)}\label{eq:DEshortform}\end{equation}
for each dimension $d$ of the local dynamical systems $f_{i}^{(d)}$
separately. As the equations (\ref{eq:DEshortform}) are uncoupled
for any two different dimensions $d$ and $d'$, we treat these separately
and drop the index $(d)$ from now on.

Repeated evaluations of the equations of motion (\ref{eq:general-equation-2})
of the system at different times $t_{m}$ thus comprise a simple and
implicit restriction on the network topology $J_{ij}$ as follows:
writing $X_{i,m}=\dot{x}_{i,m}-f_{i,m}$, these equations constitute
the matrix equation \begin{equation}
X_{i}=J_{i}G_{i}\label{eq:matrix}\end{equation}
 where $X_{i}\in\mathbb{R}^{1\times M}$, $J_{i}\in\mathbb{R}^{1\times N}$
and $G_{i}\in\mathbb{R}^{N\times M}.$ Here the elements of the \emph{i}-th
row of $J$ are given by $J_{i}$ and comprise the sequence $(J_{ij})_{j\in\{1,\ldots,N\}}$
of all input coupling strengths to unit \emph{i}. 

Can we rewrite this equation explicitly for $J_{i}$? Generically,
$M>N$, and we wish to solve this overdetermined problem by minimising
the error function given by \begin{equation}
E_{i}\left(\hat{J}_{i}\right)=\sum_{m=1}^{M}\left(x_{im}-\sum_{k=1}^{N}\hat{J}_{ik}g_{ikm}\right)^{2}\end{equation}
for the best (in Euclidean ($\ell_{2}$) norm) solution $\hat{J}_{i}$.
Here $\hat{J}_{ik}$ represents the reconstructed value of the real
coupling strength $J_{ik}$. Equating to zero the derivatives of the
error function with respect to the matrix elements, $\frac{\partial}{\partial J_{ik}}E_{i}\left(\hat{J}_{i}\right)\stackrel{!}{=}0$,
yields an analytical solution to $\ell_{2}$ error-minimisation given
by 

\begin{equation}
\hat{J}_{i}=XG_{i}^{T}\left(G_{i}G_{i}^{T}\right)^{-1}\label{eq:analytical-sol}\end{equation}
and thus the set of input coupling strengths (and input connectivity)
of unit $i$.\emph{ }Evaluating such equations for all $i\in\{1,\ldots,N\}$
yields the complete reconstructed network $\hat{J}$. This mathematical
form of minimum $\ell_{2}$-norm solution is implemented in many mathematical
packages (e.g. as the mrdivide function in MATLAB~\cite{matlab:2006}).

\section{Performance for Different Collective Dynamics}

How does this theoretical method perform in applications on data?
To illustrate the performance of the method and its insensitivity
to the type of dynamics, we apply it to four distinct collective dynamical
states ranging from simple periodic synchronous dynamics to very complex,
highly chaotic asynchronous states. We choose networks of Rössler
oscillators that can exhibit a rich repertoire of collective dynamics
from multi-dimensional chaos to periodicity and from global synchrony
to asynchrony (see Fig.~\ref{fig:fig1}), depending on local unit
parameters and coupling functions. 

\begin{figure}
\hspace{25mm}\includegraphics[width=0.75\columnwidth]{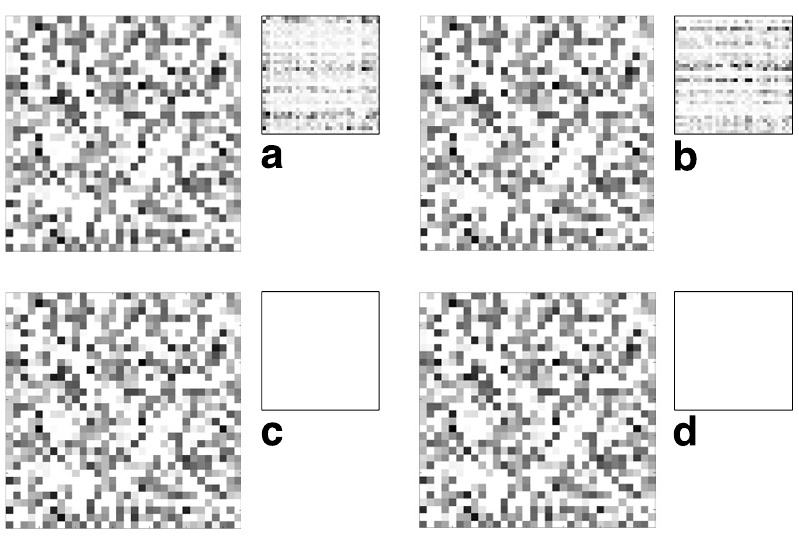}

\caption{\textsf{\textbf{\small \label{fig:reconstruction}}}Successful reconstruction
of the topology of networks from very different dynamics. Each of
the panels (a)-(d) shows a reconstruction of the network's topology
from the dynamics shown in the respective panels (a)-(d) of Figure~1.
The smaller panels show absolute differences (magnified by a factor
of $10^{8}$) from the actual network (picked randomly from an ensemble
of networks with connection probability $p=0.5$). The reconstructions
(a)-(d) used trajectories from very different dynamical states, as
shown in Figure~\ref{fig:fig1} (a)-(d). Reconstructions in (c) and
(d) have lower errors as they utilise whole dynamical trajectories,
instead of transients towards synchrony, as in (a) and (b).}

\end{figure}

\subsection{Successful reconstruction from different dynamics }

The dynamics of each Rössler oscillator~\cite{Rossler:1976p15735}
is given by the three ordinary differential equations

\[
\dot{x}_{i}=-y_{i}-z_{i}+\sum_{j=1}^{N}J_{ij}f\left(x_{i},x_{j}\right)\]

\[
\dot{y_{i}}=x_{i}+a_{i}y_{i}\]

\begin{equation}
\dot{z}_{i}=b_{i}+z_{i}(x_{i}-c_{i})\label{eq:rossler-eq}\end{equation}
where $a_{i},b_{i},c_{i}$ are local parameters. The coupling function
was set to $f\left(x_{i},x_{j}\right)=x_{j}-x_{i}$ to induce synchronisation
and to $f\left(x_{i},x_{j}\right)=\sin\left(x_{j}\right)$ to prevent
it. The local unit parameters $a_{i}$, \foreignlanguage{ngerman}{$b_{i}$}
and $c_{i}$ of the Rössler oscillators are chosen to induce either
chaotic or periodic dynamics. The parameters were treated as unknown
and are not needed for the reconstruction of the network topology.
This is the case in this example, since the equations where the parameters
of network topology occur (in the \emph{x}-dimension) do not contain
any other (unknown) parameter. We now demonstrate a reconstruction
of the network in all four dynamical paradigms illustrated in Fig.~\ref{fig:fig1}:
periodic synchronous, chaotic synchronous, periodic asynchronous and
chaotic asynchronous collective states. 

Reconstruction \emph{in praxis} works as follows: for networks exhibiting
synchronised dynamics, the coupling function $f\left(x_{i},x_{j}\right)=x_{j}-x_{i}$
is uniformly zero for all time for all units, revealing no information
about the network topology. Nevertheless, the network can still be
reconstructed from its \emph{transient} dynamics towards the synchronous
state. In general, by substituting $X_{i,m}=\dot{x}_{i}\left(t_{m}\right)+y_{i}\left(t_{m}\right)+z_{i}\left(t_{m}\right)$
and $G_{ij,m}=f\left(x_{i}(t_{m}),x_{j}(t_{m})\right)$ in (\ref{eq:matrix}),
we find the least-squares reconstructed network according to (\ref{eq:analytical-sol}).
Since each unit has at most \emph{$N-1$ }incoming links of unknown
weights, the diagonals in the reconstructed $\hat{J}$ are zero. A
reliable reconstruction of the network from trajectories (as shown
in Figure~\ref{fig:fig1}) is illustrated in Figure~\ref{fig:reconstruction},
for all four dynamical paradigms considered.

\subsection{Quality of Reconstruction}

\begin{figure}[!th]
\hspace{20mm}\includegraphics[width=0.75\columnwidth]{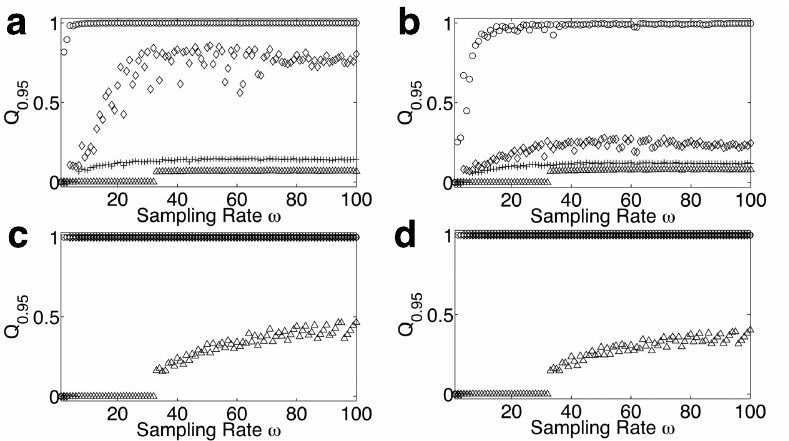}

\caption{\textsf{\textbf{\small \label{fig:QvsW}}} Quality of reconstruction
($Q_{0.95}$) as function of sampling rate $\omega$ and observation
time $T$. Good quality reconstruction can be achieved even at low
sampling rates. Underlying system dynamics used for reconstruction
in panels (a)-(d) are shown in Fig.~\ref{fig:fig1} (a)-(d). Lengths
of observations are\emph{ $T=1$} ($\triangle$), $T=5$ ($+$), $T=10$
($\diamondsuit$), and $T=20$ ($\bigcirc$). Each point is the result
of averaging over 50 networks. }

\end{figure}

How accurate is such a reconstruction? The quality of reconstruction

\begin{equation}
Q_{\alpha}:=\frac{1}{N^{2}}\sum_{i,j}H\left(\left(1-\alpha\right)-\Delta J_{ij}\right)\in\left[0,1\right]\end{equation}
is defined as the fraction of coupling strengths that are considered
correct. Here $\alpha\leq1$ is the required accuracy of the coupling
strengths and $H$ is the Heaviside step function, $H\left(x\right)=1$
for $x\geq0$ and is $H(x)=0$ otherwise. The normalized element-wise
difference between the reconstructed and real network is

\begin{equation}
\Delta J_{ij}:=\left|\hat{J_{ij}}-J_{ij}\right|/\left(2J_{max}\right)\end{equation}
where $J_{max}=\max_{i',j'}\left\{ \left|J_{i',j'}\right|,\left|\hat{J}_{i',j'}\right|\right\} $.
Typically, the quality of reconstruction increases with both the sampling
rate and the length of trajectory observed, becoming close to 1 even
at lower sampling rates for longer times (see Fig.~\ref{fig:QvsW}). 

\subsection{Required observation time}

What is the minimum length of observation required to reconstruct
a network? We define 

\begin{equation}
T_{q,\alpha,\omega}:=\min\left\{ T\left|Q_{\alpha}\left(T\right)\geq q\right.\right\} \end{equation}
to be the minimum length of time of observation at a sampling rate
$\omega$ required for accurate reconstruction at quality level \emph{$q$}
at accuracy $\alpha$. A general observation is that with increasing
sampling rate $\omega$ or increasing observation time $T$, the quality
increases, due to more accurate information that is obtained about
the system's states. In general, however, it is not only the total
number $T\times\omega$ of restricting equations (per node and per
dimension) that controls the quality. For instance, at a given observation
time, high quality close to $Q_{0.95}=1$ is reached even at lower
sampling frequency if the collective dynamics is more irregular, cf.,
Figure~\ref{fig:QvsW}a,b vs. Figure~\ref{fig:QvsW}c,d. We ascribe
this to the lower degree of correlation among observations at different
times that is required for accurate reconstruction numerics in (\ref{eq:analytical-sol}),
e.g. for irregular dynamics, sample points more distant in time provide
more relevant information increase about the system as they are less
correlated than closer-by points. 

\begin{figure}
\includegraphics[width=1\columnwidth]{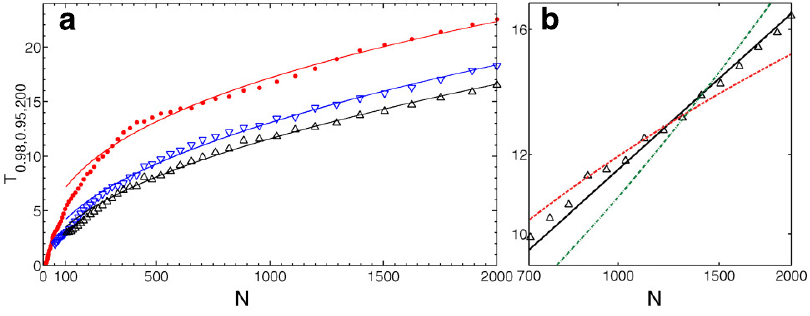}

\caption{\textsf{\textbf{\footnotesize \label{fig:T98vsN}}} Sublinear scaling
of reconstruction time with system size. (a) Minimum required time
of observation for reconstruction $\left(q=0.98,\alpha=0.95,\omega=200\right)$
grows sublinearly (presumably algebraically) with system size \emph{N
}for networks with various fixed indegrees $K=10\left(\bullet\right)$,
$K=50\left(\triangledown\right)$, $K=100\left(\vartriangle\right)$.
The fit suggests that $T_{q,\alpha,\omega}\propto N^{\gamma}$, where
the exponent of scaling $\gamma\thickapprox0.5$. (b) An algebraic
scaling (black) fits the data (\emph{K}=100, $\vartriangle$) best.
Data plotted on a bilogarithmic scale. Linear best-fits (green dashes)
overestimate and logarithmic fits (red dots) underestimate reconstruction
time. All fits to data from \emph{N}=100 to \emph{N}=2000.}

\end{figure}

How does the minimum required observation time scale with system size?
Figure~\ref{fig:T98vsN} shows $T_{0.98,0.95,200}$, the minimum
length of time of observation required to have at least $q=98\%$
of the links accurate in strength to an accuracy of at least $\alpha=0.95$,
sampled at a rate of $\omega=200$, as a function of \emph{N}. The
numerics suggest that, at fixed $\omega$, $T_{q,\alpha,\omega}$
scales sublinearly with network size\emph{ N} for reasonably small
$0<1-\alpha\ll1$ and $0<1-q\ll1$ (reasonably large $\alpha$ and
$q$). This implies that the cost of observation does not grow prohibitively
quickly, and that even large networks can be reconstructed by a single
observation of the collective dynamics.

\subsection{Robustness: Substantial noise and unknown parameters}

\begin{figure}
\includegraphics[width=1\columnwidth]{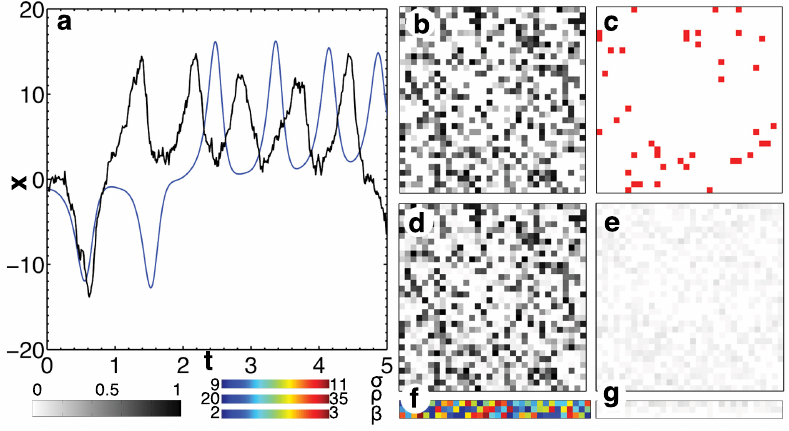}

\caption{\textsf{\textbf{\footnotesize \label{fig:fig10}}}Reconstructing a
network and unknown parameters for a system in the presence of substantial
external noise. (a) The dynamics of a unit in a network of 32 Lorenz
oscillators in the noise-free (blue) and noise-driven (black) regimes.
The network was a realisation from an ensemble of networks with edge
connection probability $p=0.5$. Starting from the same initial condition,
the noise-driven trajectory quickly deviates due to the chaotic nature
of the system. Reconstruction of the network topology (d) and parameters
(f) with corresponding absolute errors (e), (g). (b) shows the actual
network, and (c) shows the false positives in the reconstruction.
There were no false negatives. }

\end{figure}

Is reconstruction still feasible in the presence of substantial noise?
Is it possible to find unknown parameters of the local unit systems?
In the preceding examples, none of the unknown parameters appeared
in the dimension of coupling, making the problem of reconstructing
network topology distinct from that of inferring dynamical parameters.
In the following example we illustrate reconstructing networks with
intrinsic unit dynamics that are governed by arbitrary functions with
\emph{K} unknown parameters, and where the dynamics are influenced
by substantial additive noise $\xi_{i}^{\left(d\right)}$. Assuming
that all $\xi_{i}^{(d)}$ have a finite variance, we note that an
observation of the dynamics of the system yields a system of equations
linear in the \emph{K}+\emph{N} unknowns, which we can solve as before.
A simple example is a network of Lorenz oscillators~\cite{Lorenz:1963p15795},
where the dynamics of each oscillator is given by

\[
\dot{x}_{i}=\sigma_{i}\left(y_{i}-x_{i}\right)+\sum_{j=1}^{N}J_{ij}\left(x_{j}-x_{i}\right)+\eta\xi_{i}^{\left(x\right)}\]

\[
\dot{y_{i}}=x_{i}\left(\rho_{i}-z_{i}\right)-y_{i}+\eta\xi_{i}^{\left(y\right)}\]

\begin{equation}
\dot{z}_{i}=x_{i}y_{i}-\beta_{i}z_{i}+\eta\xi_{i}^{\left(z\right)}\end{equation}
where the parameters $\sigma_{i},\rho_{i}$ and $\beta_{i}$ are unknown,\textbf{
}and chosen randomly from intervals where the Lorenz system is known
to be chaotic:\textbf{ $\sigma_{i}\in\left[9,11\right],\rho_{i}\in\left[20,35\right],\beta_{i}\in\left[2,3\right]$$ $}.
The noise terms $\xi_{i}^{\left(d\right)}$, $d\in\{x,y,z\}$, $i\in\{1,\ldots,N\}$
are independent identically distributed random variables (on the discrete
simulation time scale of\textbf{ $\Delta t=0.001$}) drawn from the
standard normal distribution. The network topology \emph{and }all
parameters of the system can be reconstructed and the reconstruction
method works despite substantial interference from noise. Figure~\ref{fig:fig10}
shows a successful reconstruction of the network and of all dynamical
parameters for a network of heterogeneous Lorenz oscillators, where
the noise amplitude $\eta=0.5$ is chosen such that it drastically
alters the dynamics from its deterministic counterpart (black vs.
blue curve in Figure 5a). This illustrates by example that the theory
is insensitive to additive noise and capable of successful reconstruction,
as desired for generic real-world systems. 

\section{Conclusion}

In summary, we have introduced a simple robust method to infer connection
topology from observations of deterministic and noisy network dynamical
systems, where the functional form of the evolution equations is known.

Our method is unique in the following ways. A simple sufficiently
long observation of the system dynamics suffices to reconstruct the
network topology and coupling strengths. In many cases, experimental
access to the system, say, to introduce `control signals', as in~\cite{Timme:2007p14319,Yu:2008p13684}
may not be possible, and the method proposed here does not require
any form of system intervention. Furthermore, the type of collective
dynamics is not restricted to, e.g., fixed points, periodic orbits,
synchronous states, or any other specific type of motion, cf.~\cite{Gardner:2003p9614,Timme:2007p14319}.
Using entire dynamical trajectories, including transients, for the
reconstruction, we demonstrate robust reconstruction for a wide range
of observed dynamical states, from asynchronous chaotic states to
transient states towards global synchrony. Moreover, we treated the
system as a `gray box', where we know some general principles of the
system (like the coupling functions) but lack the details, like network
structure or intrinsic parameters. Given (e.g. experimental) dynamical
observation data, our theory provides an explicit analytic solution
to the inverse problem of finding the network structure. This solution
is a direct restatement of the differential equations governing the
dynamics of the system, and is thus conceptually the simplest possible.
This simplicity suggests highest attainable quality for such inverse
problems. 

The method scales sublinearly with network size, seems robust against
substantial addition of noise, and thus provides a promising complement
to existing reconstruction methods. Thus, our method offers a conceptual
simplification over other methods that make the same assumptions we
do, but rely on more complex techniques like copy-synchronisation
(called auto-synchronization in~\cite{Yu:2008p13684}) or the use
of topology estimating clone models or control signals~\cite{Yu:2006p13046}.
Further, the method suggested here is capable of reconstructing network
structure from a simple observation of the system's dynamics, without
resorting to any external intervention to drive the system into or
from some canonical state, as in~\cite{Timme:2007p14319}. 

Efforts to understand the general interplay of network structure and
dynamics have yielded several promising approaches, mainly applicable
to smaller systems. Notable among the forward methods, i.e., methods
that predict dynamical features from knowledge of network topology
are those that study the propagation of a harmonic perturbation through
a network of coupled phase oscillators~\cite{Zanette:2004p9940,Kori:2004p10015},
and methods to predict disordered dynamics from structures of strongly
connected components in the network~\cite{Timme:2006p987}. Cimponeriu
et al.~\cite{Cimponeriu:2004p14049} introduce two methods to estimate
the interaction delay in weak coupling between two self-sustained
oscillators from observed dynamical time series. Arenas et al.~\cite{Arenas:2006p10049}
show that times of synchronisation can reveal the hierarchical structure
of a network, revealing a connection between synchronisation dynamics
and topological clustering. In an alternative approach, Memmesheimer
and Timme~\cite{Memmesheimer:2006p10092,Memmesheimer:2006p10093}
present an analytical method to design networks of spiking neurons
that display a required spike pattern. Other inverse methods have
relied on stochastic optimisation~\cite{Makarov:2005p9589} to fit
a model of a network of spiking neurons to an observation of a real
network to infer its topological parameters. 

Several avenues for further research present themselves. As suggested
by previous work~\cite{Timme:2007p14319,Tegner:2003p9754}, minimising
the $\ell_{1}$ norm, instead of the $\ell_{2}$ norm as used here
may result in more efficient reconstruction of sparse networks~\cite{Napoletani:2008p13092}.
In addition, our preliminary studies suggest that this highly overdetermined
inverse problem can be reduced to an exactly-determined problem by
selectively choosing points on the time series to ensure that the
resultant system of equations is maximally linearly independent. This
reduction significantly reduces the cost of computation to reconstruct
large networks. Furthermore, it is straightforward to extend this
method to coupled map networks and to systems with delay. Finally,
recent studies show that a method analogous to the one presented here
for smoothly coupled systems is capable of reconstructing networks
of pulse-coupled systems such as integrate-and-fire neurons~\cite{VanBussel:2010p14624}.

\ack{}{We thank C. Kirst for valuable discussions and advice. We thank the
Federal Ministry of Education and Research (BMBF) Germany for support
under grant number 01GQ0430 to the Bernstein Center for Computational
Neuroscience (BCCN), Göttingen.}

\section*{References}{}

\bibliographystyle{iopart-num}
\bibliography{intfcd}

\end{document}